\documentstyle[twoside,fleqn,espcrc2,epsfig]{article}
\def\f {\beta_f}
\def\a {\beta_a}
\def\v {\beta_v}
\def\b {\beta}

\def\l {\langle L_a \rangle}

\def\fa {4^3 \times 4}
\def\sa {6^3 \times 4}
\def\ea {8^3 \times 4}
\def\fs {4^4}
\def\xs {6^4}
\def\es {8^4}
\def\tw {8^3 \times 2}

\newcommand{\AmS}{{\protect\the\textfont2
  A\kern-.1667em\lower.5ex\hbox{M}\kern-.125emS}}

\begin{document}
\title{Phase Diagram of SO(3) Lattice Gauge Theory at Finite 
Temperature}
\author{\vspace*{-2.0cm}
\hfill \mbox{TIFR/TH/97-43} \\
	\vspace*{2.0cm}        
        Saumen Datta$^{\rm a}$ and Rajiv V. Gavai\address{Theory Group, Tata 
        Institute of Fundamental Research, \\
        Homi Bhabha Road, Mumbai 400 005, India}%
         \thanks{Speaker at the symposium}}
       
\begin{abstract}

The phase diagram of $SO(3)$ lattice gauge theory at finite
temperature is investigated by Monte Carlo techniques with a view i) to
understand the relationship between the deconfinement phase transitions in
the $SU(2)$ and $SO(3)$ lattice gauge theories and ii) to resolve the
current ambiguity of the nature  of the high temperature phases of the 
latter. Phases with positive and negative adjoint Polyakov loop,
$L_a$, are shown to have the same physics. A first order deconfining
phase transition is found for $N_t = 4$.

\end{abstract}

\maketitle

\section{Introduction}

Since the continuum limit of a lattice gauge theory is 
governed by its 2-loop $\beta$-function, one expects 
the physics of confinement and deconfinement 
for pure $SU(2)$ gauge theory to be identical to that of pure $SO(3)$ 
gauge theory.  On the other hand, $SO(3)$ does not have the $Z(2)$ center 
symmetry whose spontaneous breakdown in the case of the $SU(2)$ theory
indicates its deconfinement transition.  This makes the investigation of
the phase diagram of the $SO(3)$ gauge theory especially interesting and
important.  It has been argued\cite{smi} that the deconfinement transition 
for the $SO(3)$ lattice gauge theory may show up
as a cross over which sharpens in the continuum limit to give an
Ising-like second order phase transition.

Another reason for investigating the finite temperature transition in
$SO(3)$ gauge theory is that it is supposed\cite{bha} to have a bulk phase
transition and may thus provide a test case for studying the interplay
between these different types of phase transitions.  Recently, 
simulations of the
Bhanot-Creutz action for SU(2) gauge theory\cite{bha},
\begin{equation}
S=\sum_p \bigl\lbrace \f (1 - {1 \over 2} {\rm Tr}_f ~ U_p) + \a (1 -
{1 \over 3} {\rm Tr}_a
 ~ U_p) \bigr\rbrace, \label{BHA}
\end{equation}
at finite temperature revealed\cite{gav1} that the known deconfinement 
transition point in usual Wilson action becomes a line in the $\f$-$\a$
plane and joins the bulk transition line seen in \cite{bha}. The order
of the deconfinement transition was also seen to change from second to
first for $ \a \ge 1.25 $. The studies in \cite{gav1} were all done for a 
relatively small $\a$, i.e., close to the Wilson action.  $SO(3)$ gauge theory
provides an example far away from it.

A study of finite temperature SO(3) gauge 
theory was carried out in \cite{sri} and a deconfining transition for
this theory was found.  However, there was some ambiguity about the nature 
of the high temperature phase and the order of the phase transition
in \cite{sri}. In this work, we attempt to clarify these
ambiguities. 

\section{Actions and Observables}

The Wilson action for SO(3) gauge theory is
\begin{equation}
S = \b \sum_p (1 - {1 \over 3} {\rm Tr} ~ U_p)~~, \label{ADJ}
\end{equation}
where $U_p$ denotes the directed product of the link variables,
$U_\mu (x) \in SO(3)$, around an elementary plaquette p. The action 
(\ref{BHA}) for $\f $=0 also corresponds to an
$SO(3)$ gauge theory which was found in \cite{bha} to have a first 
order bulk transition at $\a \sim 2.5$.  A  third action 
we used is the Halliday-Schwimmer action \cite{hal}
\begin{equation}
S=\v \sum_p (1 - {1 \over 2}  \sigma_p {\rm Tr}_f ~ U_p) ~~.~~\label{HAL}
\end{equation}
Here the link variables $U_\mu (x) \in SU(2)$
and $\sigma_p = \pm 1$. Besides the integration over the link variables, 
the partition function in this case also contains a summation over
all possible values of $\lbrace \sigma_p \rbrace$.
It too shows \cite{hal} a first order bulk phase transition at $\v \sim
4.5$. The chief advantage of this action is that both the link variables
$U_\mu$ and $\sigma_p$ can be updated using heat-bath algorithms.

We studied the adjoint plaquette $P$, defined as the 
average of ${1 \over 3} {\rm Tr}_a U_p$ [$\sigma_p{\rm Tr}_f U_p$]
over all plaquettes for actions (\ref{BHA}) and (\ref{ADJ}) 
[ action (\ref{HAL})]. 
We also measured $\l$, which is the average over all spatial sites of the 
adjoint Polyakov loop, defined by 
$L_a (\vec r) = \rm{Tr}_a ~ \prod_{i=1}^{N_t} U_t (\vec r, i)$.
Note that $\l$ is not an order parameter. 
Since $\l$ can be thought of as a measure of the free energy of 
an adjoint quark which can be screened by gluons created from
the vacuum, it is not constrained to be zero in the confined phase. 
Similarly an adjoint Wilson loop is not supposed to show area law
in the confining phase. However, 
creation of gluon pairs from vacuum costs a considerable amount of
energy as glueballs are heavy. It may therefore be favourable for adjoint 
quarks also to have a string between them, at least when they are not
too far separated.  
Using the action (\ref{BHA}) for $\f=0$ on a $7^3 \times 3$ lattice, Ref.
\cite{sri} found that $\l$ was consistent with zero till $\a \sim 2.5$, 
after which it became nonzero, indicating a deconfinement transition 
around this value of $\a$.

\section{The High Temperature Phase}

An unexpected and curious result of Ref. \cite{sri} was that after
becoming nonzero in the high temperature phase, $\l$ 
settles into either a positive value ($\to 3$ as $\a \to \infty$), or 
a negative value ($\to -1$ as $\a \to \infty$), the two states being
degenerate in free energy. In \cite{sri} the negative $\l$ state was
interpreted as the manifestation of another zero temperature confined phase. 

\begin{figure}[htbp]\begin{center}
\epsfig{height=73mm,width=50mm,angle=-90,file=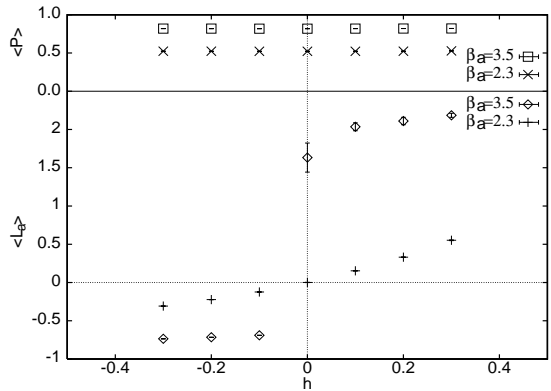}
\caption{$\langle P \rangle$ and $\l$ 
as a function of magnetic field $h$.} 
\end{center}\end{figure}

	We have carried out a number of tests in order to understand
the nature of the negative $\l$ state. First, it was checked that the
appearance of this phase is not due to any algorithmic problem by
observing that it appears for all the three actions discussed above.
We then checked that the value of $\l$ is
quite stable against changes in spatial lattice sizes from
$9^3 \times 3$ to $18^3 \times 3$ lattices.
Next, we introduced a polarising `magnetic field' by adding a term 
$h \sum_{\vec x} L_a(\vec x)$ to the action.  As shown in the upper
half of Fig. 1, 
the average plaquette P on a $7^3 \times 3$ lattice is not affected 
strongly by this term either below the transition ($\a=2.3$) or above 
the transition ($\a=3.5$).  However, $\l$ is.  Irrespective of the start,
it converges to a unique value whose sign is determined by that of $h$.
This is
similar to the $SU(2)$ case and suggests strongly that the high temperature 
phase of the $SO(3)$ gauge theory also manifests itself in two ways 
corresponding to positive and negative $\l$.  Also note in Fig. 1 that the 
extrapolation of $\l$ to vanishing $h$ yields a value consistent with zero 
below the phase transition.

A further test of whether the physics of these two phases is the same is
the equality of the correlation lengths in these phases.  We measured the
correlation function, defined by 
\begin{equation}
\Gamma(r) = \sum_i \sum_{\vec x} \langle L_a(\vec x + r e_i) ~  L_a(\vec
x) \rangle~~,~~
\end{equation}
on an $\ea$ lattice for $\a = 2.3$ and for the positive and negative 
$\l$ states at $\a = 2.6$ and 3.5.
It was found that i) at $\a=2.3$, which is below the transition, the
correlator vanishes rapidly with $r$, ii) it approaches a constant above
the phase transition and iii) the constant is bigger for larger $\a$ and
bigger in the positive $\l$-phase for the same $\a$.  
The mass gap, obtained from the connected parts of the
correlator above or from their zero momentum projected versions,
was similar for {\em both} the positive and negative $\l$ states 
corresponding to both $\a = 2.6$ and 3.5, as expected for states with 
same physics.  It is, however, considerably different for $\a = 2.3$. 

\section{Order and Nature of the Transition}

In simulations
on $\fa$, $\sa$ and $\ea$ lattices with the actions (\ref{BHA}) and 
(\ref{HAL}), long metastable states were observed on all lattices near 
the transition region, signaling a possible first order transition. 
$\l$ was seen to tunnel between 
all the three states, two of which correspond to the same value of the 
action.  Runs on smaller lattices
show more tunnellings and larger fluctuations in the
positive $L_a$-phase. The estimated transition points for $\fa$, $\sa$ and 
$\ea$ lattices are $\beta_{vc} = 4.43 \pm 0.02$, $4.45 \pm 0.01$ and 
$4.45 \pm 0.01$ respectively.

Fig. 2 displays distributions of $L_a$
from the runs made at the critical couplings but from different starts. 
We performed about 100K-400K heat-bath sweeps depending on the size of the
lattice.  While the frequent tunnelling smoothens the peak structure for 
the $\fa$ lattice considerably, a clear three-peak structure is seen for 
both the $\sa$ and the $\ea$ lattices. The stability of these peaks
under changes in spatial volume  suggests the
phase transition to be of first order.  The estimates of the 
discontinuities in the plaquette, $\l_+$ and $\l_-$ are $0.0575 \pm 0.0030$,
$0.87 \pm 0.04$ and $0.28 \pm 0.04$ respectively.
It is also interesting to note that i) the
peak for the confined phase is almost precisely at zero and ii)
normalising by the maximum allowed $\l$ in each phase, the
discontinuties for both the positive and negative phases are equal,
being $0.29 \pm 0.01$ and $0.28 \pm 0.04$ respectively.

\begin{figure}[htbp]\begin{center}
\epsfig{height=73mm,width=50mm,angle=-90,file=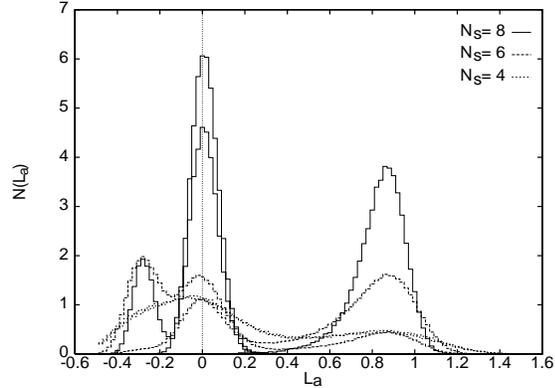}
\caption{The distribution of $L_a$ on $N_s^3 \times 4$ lattices
at their critical couplings.}
\end{center}\end{figure}

We also studied the theory on 
$\tw$, $\fs$, $\xs$ and $\es$ lattices.
On all these lattices, only one transition point was found, where
both the plaquette and $\l$ show a discontinuity. 
A clear shift in $\b_c$ was found in going from $N_t = 2$ to
$N_t = 4$ but no perceptible change in $\b_c$ was found 
in going from $N_t = 4$ to 6 and 8 for both actions (\ref{BHA}) and 
(\ref{HAL}).
This is in sharp contrast to the SU(2) case, and is also unexpected 
for a deconfinement transition. 

\section{Summary}

Our simulations with a variety of actions showed
the negative $\l$-state to be present for all
of them.  However, using a `magnetic field' term to polarise, we found
a unique $\l$ state depending on the sign of the field.
The correlation function measurements in both the phases of positive and 
negative $\l$ indicated that the two states are physically
identical high temperature deconfined phases
of SO(3) gauge theory. Although a shift in $\beta_c$ was observed in
changing $N_t$ from to 2 to 4, no further shift was seen for $N_t$ =6 
and 8 which is characteristic of a bulk phase transition.

\end{document}